\documentclass{aims_LENCOS}
\usepackage{amsmath}
\usepackage{amsfonts}
\usepackage{amssymb}
\usepackage{paralist}
\usepackage{graphics} 
\usepackage{epsfig} 
\usepackage[colorlinks=true]{hyperref}
\hypersetup{urlcolor=blue, citecolor=red}

\def\currentvolume{X}
 \def\currentissue{X}
  \def\currentyear{200X}
   \def\currentmonth{XX}
    \def\ppages{X--XX}

\newcommand{\wb}{\omega_{\mathrm{b}}}

\newcommand{\singlefig}{.5\textwidth}
\newcommand{\triplefig}{.3\textwidth}

\title[Breather--interstitial interaction]
      {Interaction of moving discrete breathers with interstitial defects}

\author[Cuevas, S\'anchez-Rey, Eilbeck and Russell]{}


 \email{jcuevas@us.es}
 \email{bernardo@us.es}
 \email{J.C.Eilbeck@hw.ac.uk}
 \email{Mica2mike@aol.com}

\begin{document}

\maketitle

\centerline{\scshape J. Cuevas, B. S\'anchez--Rey }
\medskip
{\footnotesize
 \centerline{Grupo de F\'{\i}sica No Lineal. Departamento de F\'{\i}sica Aplicada I.}
   \centerline{Escuela Universitaria Polit\'{e}cnica. Universidad de Sevilla.}
   \centerline{C/ Virgen de \'{A}frica, 7. 41011 Sevilla, Spain}
} 

\medskip

\centerline{\scshape J.C. Eilbeck and F.M. Russell}
\medskip
{\footnotesize
 \centerline{Department of Mathematics and the Maxwell Institute for Mathematical Sciences}
   \centerline{Heriot-Watt University - Riccarton, Edinburgh, EH14 4AS, UK}
}

\bigskip


\begin{abstract}
In this paper, interstitial migration generated by scattering with
a mobile breather is investigated numerically in a Frenkel-Kontorova
one-dimensional lattice.  Consistent with experimental results it is
shown that interstitial diffusion is more likely and faster than
vacancy diffusion. Our simulations support the hypothesis that a
long-range energy transport mechanism involving moving nonlinear
vibrational excitations may significantly enhance the mobility of
point defects in a crystal lattice.
\end{abstract}

\section{Introduction}

The Frenkel--Kontorova (FK) model, introduced almost 70 years ago
\cite{FK38}, is one of the most paradigmatic nonlinear systems,
whose dynamics has been widely studied during the last decades (see
\cite{BK98,BK98b,FM96,DP06} and references therein). From the point of view of
condensed matter physics, its paramount importance relies on the
ability to describe a vast number of phenomena, including different
kinds of defects such as  vacancies (Schottky defects) and, to some
extent, interstitials (Frenkel defects), which can play an important
role in the design of new materials \cite{Wuttig}.

As the FK model is basically a one-dimensional lattice of particles
subjected to a  nonlinear periodic substrate potential and a
nearest-neighbour interaction, it contains the basic ingredients to
sustain localized excitations such as topological solitons (kinks or
antikinks) or breathers. Discrete breathers (DBs), also called
intrinsic localized modes (for a very recent review about their
properties, existence proofs, computational methods and applications
see \cite{Flach08}), are exact solutions of the dynamical equations
whose energy, in contrast with normal extended wave excitations, is
not shared among lattice components but extends only over a few
lattice sites. In this sense, their spatial profiles resembles
localized vibrational modes induced by a defect site in a harmonic
lattice \cite{Sievers75}. However DBs arise only thanks to the
interplay between nonlinearity and discreteness and, for that
reason, they may occur anywhere in the lattice given sufficient
vibrational amplitude. They are also rather universal since they are
not specific to Hamiltonians with a particular form and can be found
in lattices of arbitrary dimensions. Moreover, theoretical studies
have shown DBs are linearly stable \cite{Aubry}, which implies they
can persist over very long times on top of a thermalized background
\cite{Ivanchenko}. Their investigation is not restricted to simple
toy models. Apart from indirect spectroscopic observations
\cite{Swanson}, DBs have been detected and  studied experimentally
in such different macroscopic systems as waveguide arrays
\cite{Eisenberg}, micromechanical cantilevers \cite{Sievers06},
antiferromagnetic structures \cite{Sato} and Josephson-junctions
\cite{Trias}.

In this context, an interesting problem that has attracted much
attention in recent years is the interaction between a moving
localized excitation and a lattice defect. The problem has been
addressed within different frameworks: impurities \cite{CPAR02,FPM94},
lattice junctions \cite{BSS02,ARACL06}, bending points of a polymer chain
\cite{TSI02,CK04,LCBAG04}, but most studies assume that the position of the
defect is fixed and is not able to move along the lattice. Of
current interest is the interaction between lattice defects and
moving localized excitations, which might result in movement of the
defect. This is especially true in the case of interactions arising
during irradiation of solids by swift particles, which usually
involve the creation of DBs of either longitudinal or transverse
optical mode type.

    The possibility of such interactions arose in the study of high
energy charged particles passing through crystals of muscovite, when
scattering events were postulated to create many moving highly
energy DBs. It was suggested that when such  DBs (there called
quodons) reached the end of a chain, which represents a defect in a
chain, it might cause the last atom to be ejected from the surface
\cite{RC95}. This prediction was supported by studies using both
mechanical and numerical models \cite{MER98}. Subsequently, it was
verified by experiment using a natural crystal of muscovite \cite{RE07}.
In the experiment one edge
of a crystal was bombarded with alpha particles at near grazing
incidence to create moving DBs. These propagated in chain directions
in the layered crystal and caused a proportionate ejection of atoms
from a remote edge of the crystal that was $>10^7$ unit cells
distance in a chain direction from the site of bombardment. As this
experiment was performed at 300K it not only verified the prediction
but also demonstrated the stability of these mobile DBs against
thermal motion.

Other irradiation studies have provided more empirical signs for the
interaction of DBs with defects. For
instance, in ref.\ \cite{SAR00} the authors provide evidence that,
after irradiating a silicon crystal with silver ions, a pileup of
lattice defects is accomplished at locations spatially separated
from the irradiation site. The evidence indicated that defects could
be swept by up to about 1 micron from the irradiated region. This
effect was ascribed to the propagation of highly localized packets
of vibrational energy, or DBs, created by the bombardment of heavy
ions.

Another ion-induced, athermal transport process was reported in
ref.\ \cite{AMM06}. In this case
interstitial N diffusion in austenitic stainless steel under Ar ion
bombardment was investigated. It was found that N mobility increases
in depths several orders of magnitude larger than the ion
penetration depth. This irradiation-induced enhancement of N
diffusion is consistent with previous observations which show a
dependence of the nitriding depth on ion energy \cite{WDWVWM97} and
also on the crystalline orientation \cite{ARTDPM05},  but no
conventional mechanism of diffusion can explain them. For this
reason it was suggested that diffusion of interstitial atoms  might
be assisted by highly anharmonic localized excitations which
propagate distances well beyond the ion penetration depth.

    Interstitial atoms reside in potential wells between the lattice
atoms. When a breather propagates it strongly disturbs the lattice
locally. If it passes near an interstitial these oscillatory motions
will distort the potential well confining the interstitial and will
affect significantly its mobility.  Interstitial motion consists of
jumps from one potential well to the next. Since experimental
measures deal with concentration depth profiles, interstitial
diffusion process  can be analyzed in terms of an effective movement
along a one-dimensional chain of potential wells. Moreover the
presence of an interstitial modifies potentials in adjacent atomic
chains, causing the spacing between the two nearest atoms in a chain
to the interstitial to increase. Therefore, in a first
approximation,  an interstitial  can be modelled introducing and
additional particle in a one-dimensional system and this provides
the link to the FK model.

In this paper, using a FK model with nonlinear nearest-neighbour
interaction, it is shown that migration of the disturbance in a
chain caused by an interstitial can be induced by scattering with a
mobile longitudinal mode breather. Comparison with previous work on
vacancies migration \cite{CKAER03,CASR06} also suggests that,
according to experimental results, interstitial mobility is more
likely and faster than that of vacancy defects. Of course, the
specific constraints of a one-dimensional system implies that care
is needed when attempting to carry over results to higher
dimensional lattices. Nevertheless we think that a one-dimensional
study is a necessary and useful first step before approaching the
problem with a more realistic and complex two or three-dimensional
model.

\section{The model}
    As described in the introduction, the F-K model  consists
of a chain of interacting particles subject to a periodic
substrate potential. This system is described by the following
Hamiltonian:
\begin{equation}
    H=\sum_{n=1}^N\frac{1}{2} m\dot x_n^2+V(x_n)+ W(x_n-x_{n-1}) \quad ,
\end{equation}
where $x_n$ is the absolute coordinate of the $n$-th particle. The
corresponding dynamical equations are
\begin{equation}\label{eq:dyn}
    m \ddot x_n+ V'(x_n)+[W'(x_n-x_{n-1})-W'(x_{n+1}-x_n)]=0, \quad
    n\in \mathbb{Z}.
\end{equation}

    In order to investigate interstitial mobility we have chosen a
cosine potential with the lattice period $a$
\begin{equation}
    V(x)=\frac{a^2}{4\pi^2}[1-\cos(2\pi x/a)]\; ,
\end{equation}
as the simplest, periodic substrate potential, with the linear
frequency normalized to unity $\omega_0=\sqrt{V''(0)}=1$.

For the interaction between particles, we have selected the Morse
potential
\begin{equation}
    W(x)=\frac{C}{2b^2} [e^{-b(x-a)}-1]^2, x>0.
\end{equation}
which has a minimum at the lattice period $a$ and a hard part that
prevents particles from crossing. The well depth of this potential
is $C/2b^2$ while $b^{-1}$ is a measure of the well width. Its
curvature at the bottom is given by $C=W''(a)$, so that we can
modulate the strength of the interaction  without changing its
curvature by varying parameter $b$.

\begin{figure}
\begin{center}
\begin{tabular}{cc}
    \includegraphics[width=\singlefig]{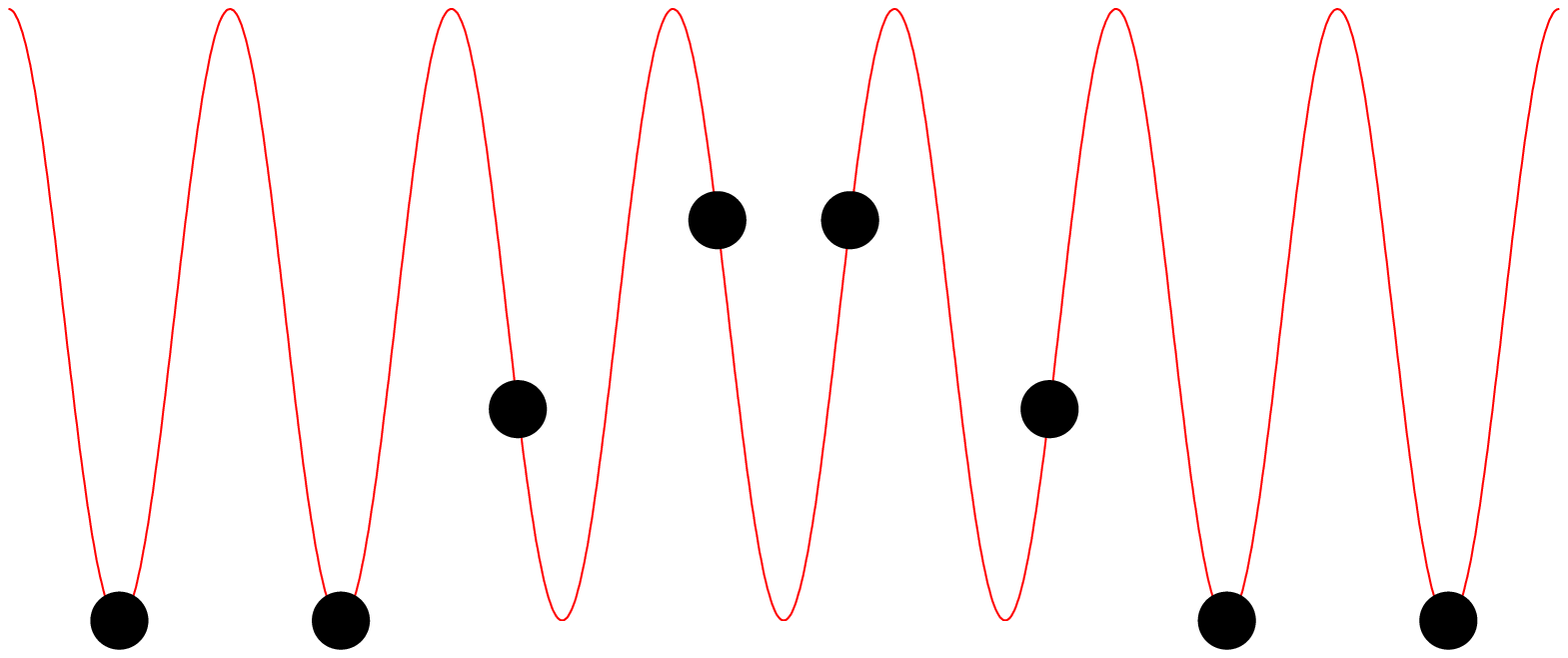} &
     \includegraphics[width=\singlefig]{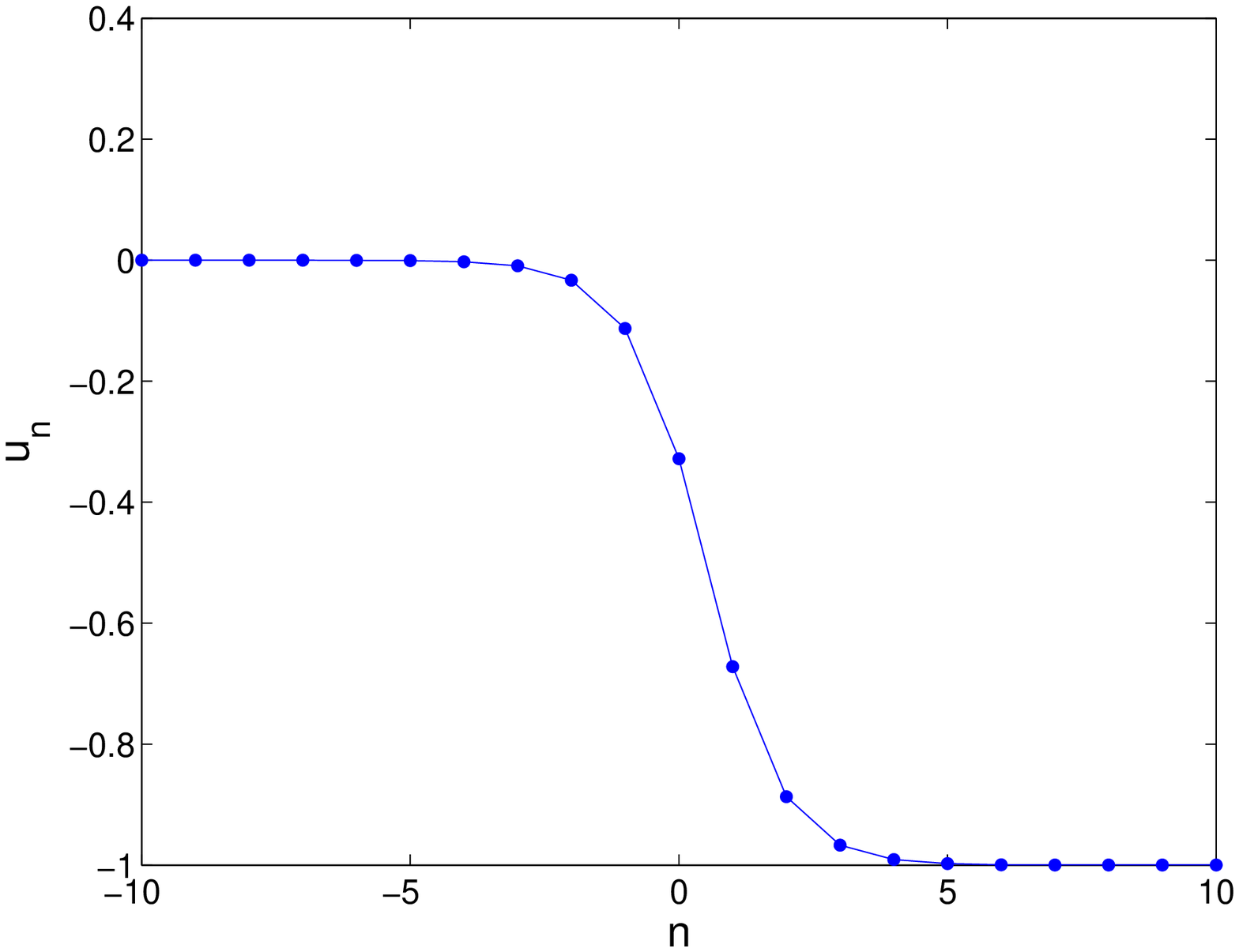}\\
     \includegraphics[width=\singlefig]{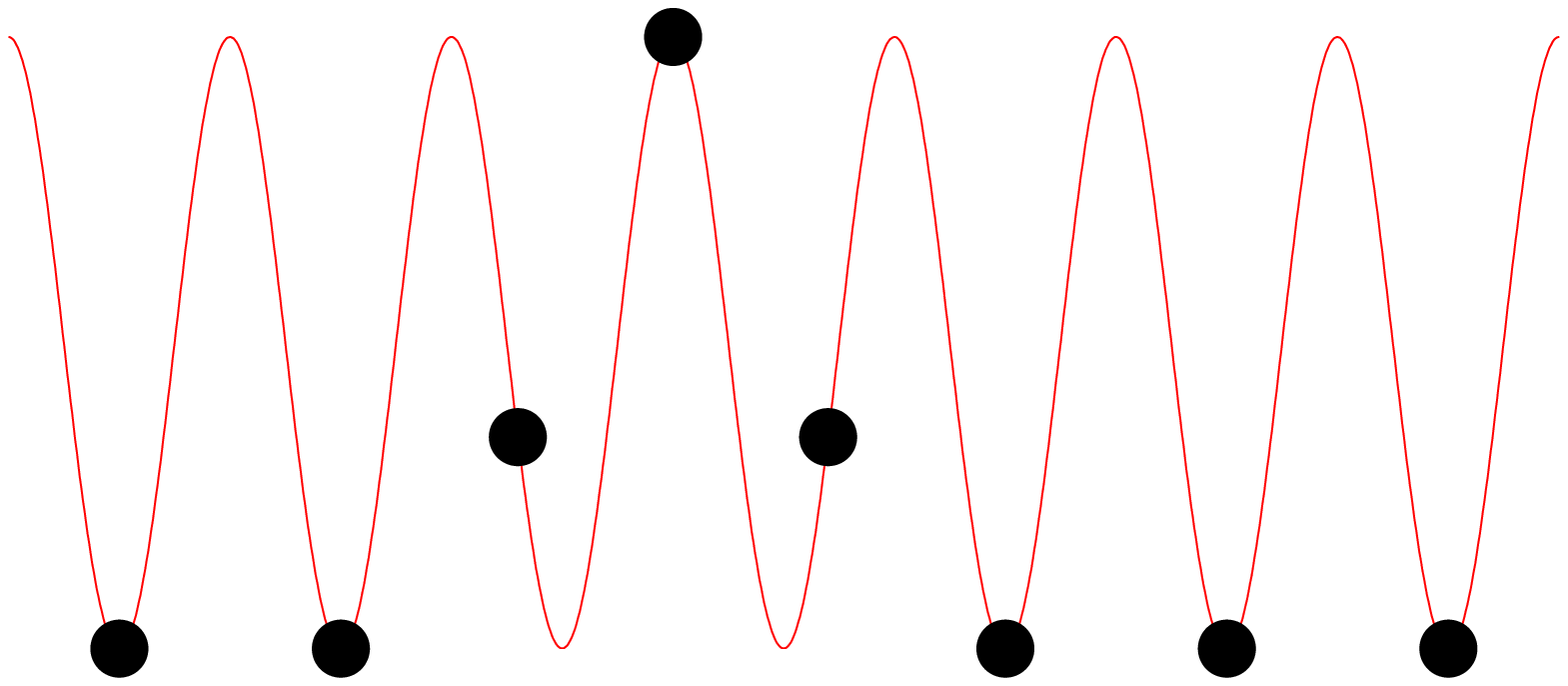} &
     \includegraphics[width=\singlefig]{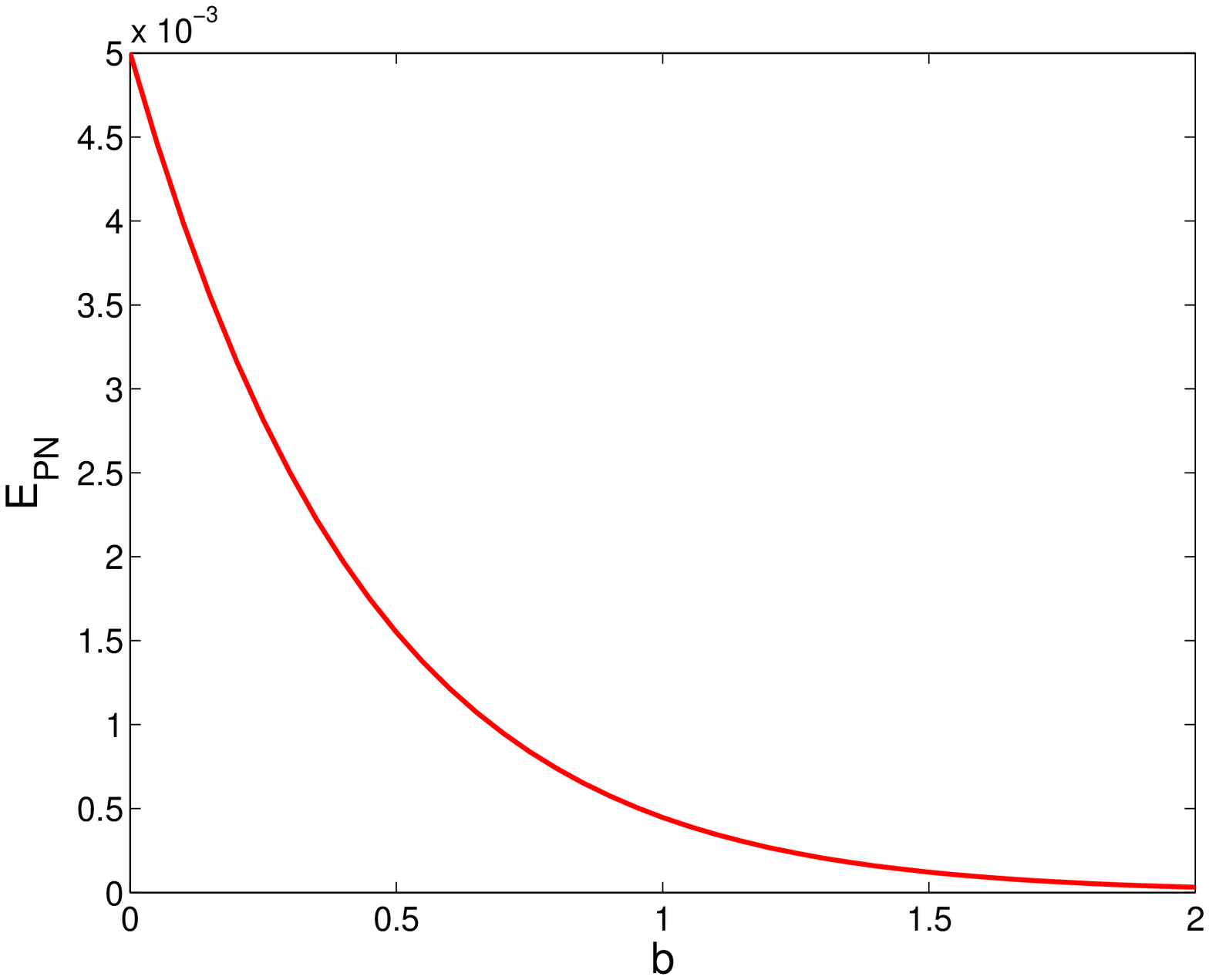}
\end{tabular}
\caption{(a) Scheme of the stable equilibrium state of the
Frenkel--Kontorova model with cosine substrate potential and Morse
nearest neighbor interaction. The doubly-occupied well represents an
interstitial. (b) Antikink corresponding to the stable equilibrium
configuration in relative coordinates for $b=1$ and $C=0.5$. (c)
Unstable equilibrium configuration. (d) Peierls-Nabarro barrier for
the antikink.} \label{fig:FK}
\end{center}
\end{figure}

In this system, an interstitial atom is represented by a
doubly occupied well of the periodic potential (see the stable
equilibrium configuration in panel (a) of Fig.\ref{fig:FK}). The
relative coordinate of each particle with respect to its equilibrium
position can be written as $u_n=x_n-na$. Using these relative
coordinates, the interstitial can be visualized as an antikink
\cite{BK98,BK98b}\footnote{Notice that in Ref. \cite{BK98b} the terms kink
  and antikink are interchanged.} as it is shown in panel (b) of
Fig.~\ref{fig:FK}. It is well-known that an antikink can be put into
movement as soon as an energy barrier, the so-called Peierls-Nabarro
barrier (PNB), is overcome. The PNB can be calculated as the energy
difference between the  unstable and stable antikink equilibrium
configurations (panels (c) and (a) of Fig.~\ref{fig:FK} respectively)
and decreases monotonically with $b$ (see panel (d)).

It is worth noting that a vacancy can be visualized as a kink in
relative coordinates. Its PNB increases with the parameter $b$ and is
always higher than the PNB of an interstitial, except for $b=0$ where
both activation energies coincide. This is in accordance with the
experimental fact that diffusion of interstitials  is faster than
that of vacancies, and support the idea that  it is necessary to
consider a nonlinear interaction potential in order to study
diffusion of defects, since $b=0$ represents the linear limit of the
Morse potential.

In our F-K chain, stationary discrete breathers can be numerically
obtained using the standard method of continuation from the
anticontinuous limit~\cite{ELS84,MA96}. Translational motion of
discrete breathers can be induced~\cite{CAT96,AC98} by adding a
perturbation $\vec{v}=\lambda (...,0,-1/\sqrt{2},0,1/\sqrt{2},0,...)$
to the velocities of the stationary breather, with the nonzero values
at the neighboring sites of the initial breather center. The resulting
DB kinetics is very smooth and resembles that of a classical free
particle.  Therefore, the total energy of a moving discrete breather
can be estimated as the sum of its vibrational internal energy, equal
to that of the stationary breather, plus its translational energy,
which is equal to the energy of the added perturbation
$K=\lambda^2/2$.

\section{Numerical study}

    In order to investigate interstitial mobility, we have generated
a  breather centered at site $n=-25$, relatively far from an
interstitial whose leftmost particle is located at $n=0$, and then launched that breather
towards it following the depinning method mentioned above.
Throughout the paper, we have normalized the lattice period $a$ and
masses to unity and have taken $C=0.5$ so that moving breathers
(MBs) exist in the system for a breather frequency $\omega_b=0.9$.

\begin{figure}
\begin{center}
\begin{tabular}{ccc}
    \includegraphics[width=\triplefig]{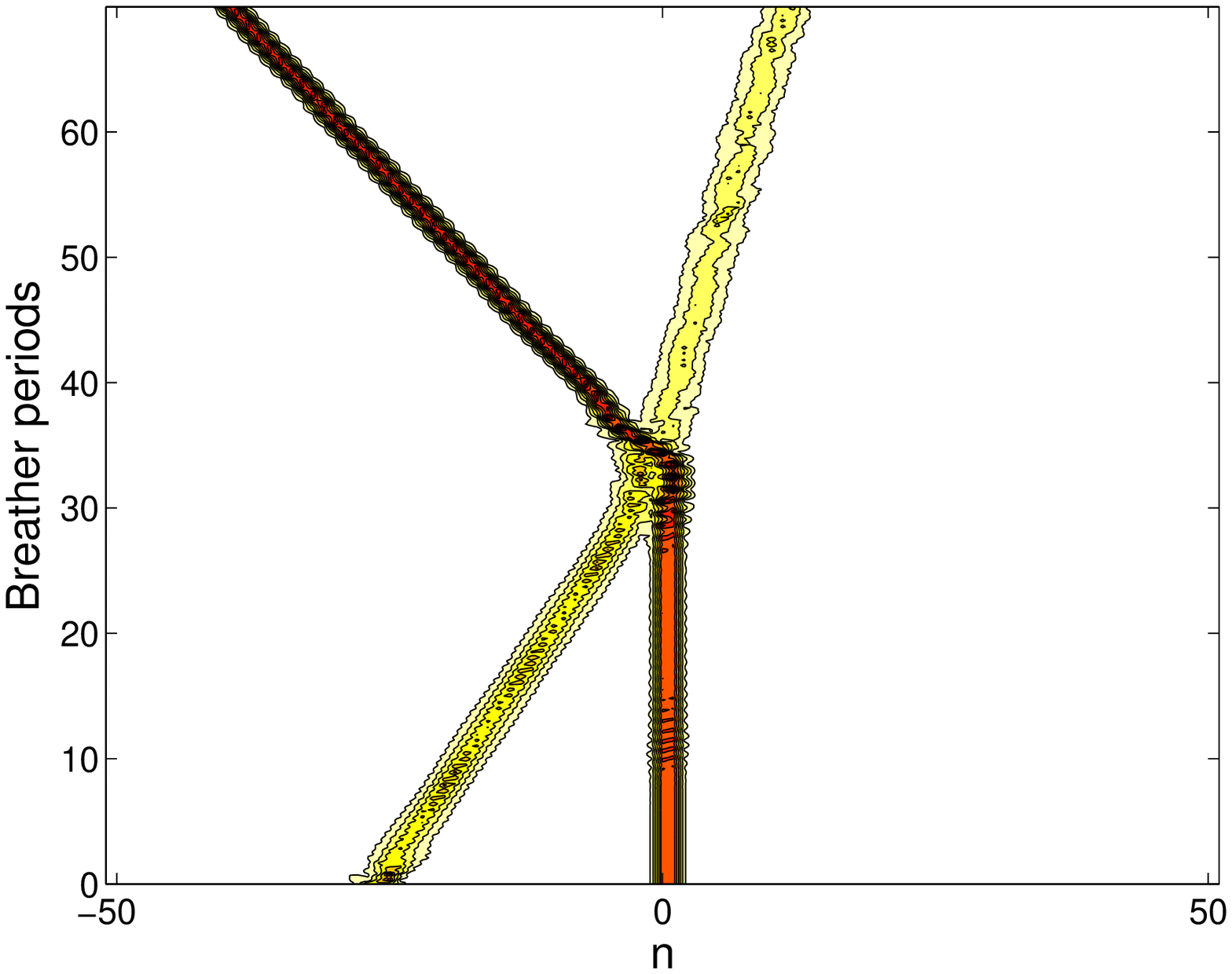} &
    \includegraphics[width=\triplefig]{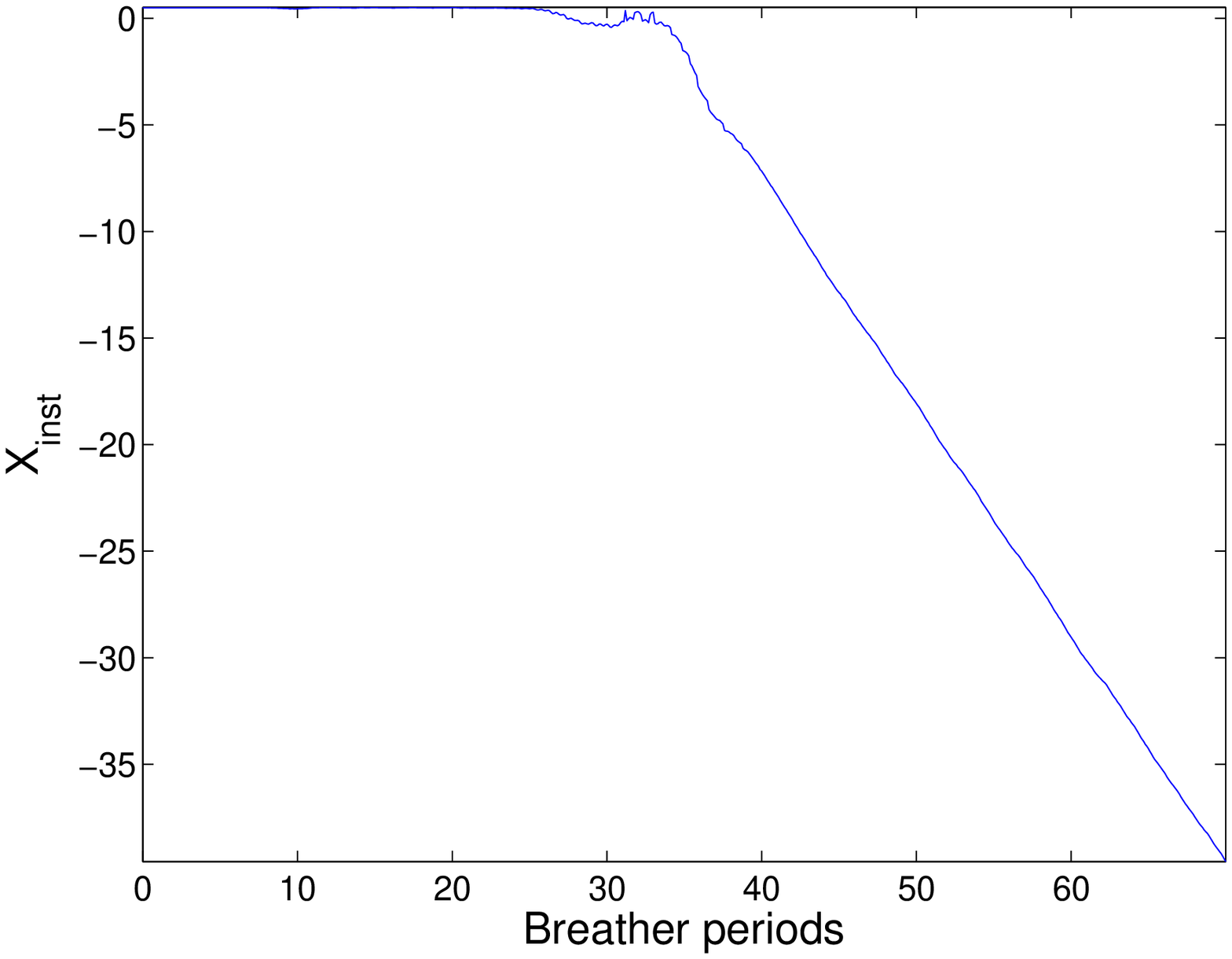} &
    \includegraphics[width=\triplefig]{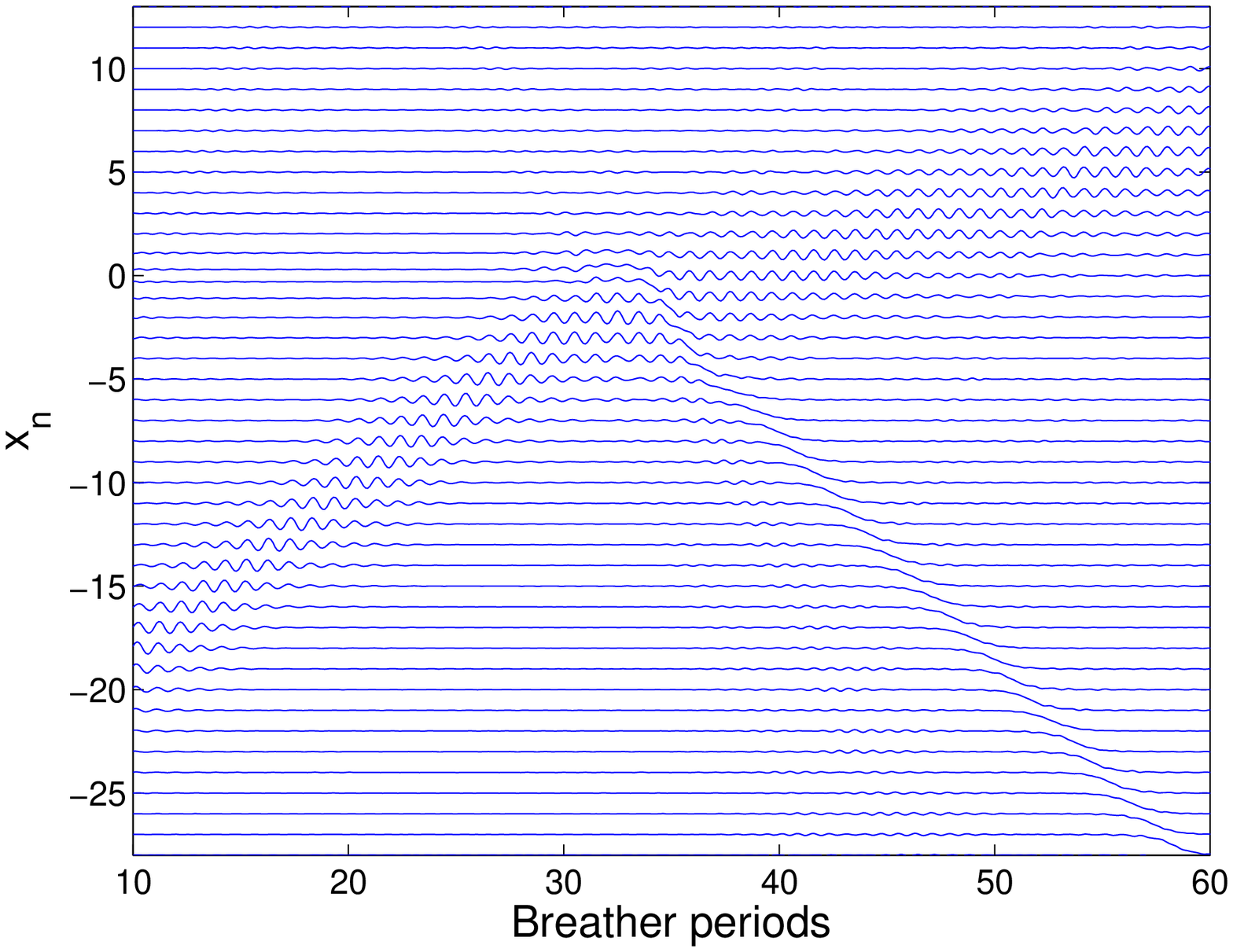} \\
\end{tabular}
\caption{(Left panel) Energy density plot, showing a backward
movement of the interstitial defect after breather scattering. The
lines join points with the same energy in time. The darker colour
the larger energy. (Central panel) Time evolution of the antikink
energy center. (Right panel) Streak plot. Parameters: $K=0.0220$ and
$b=0.5$.} \label{fig:edpbk}
\end{center}
\end{figure}

\begin{figure}
\begin{center}
\begin{tabular}{ccc}
    \includegraphics[width=\triplefig]{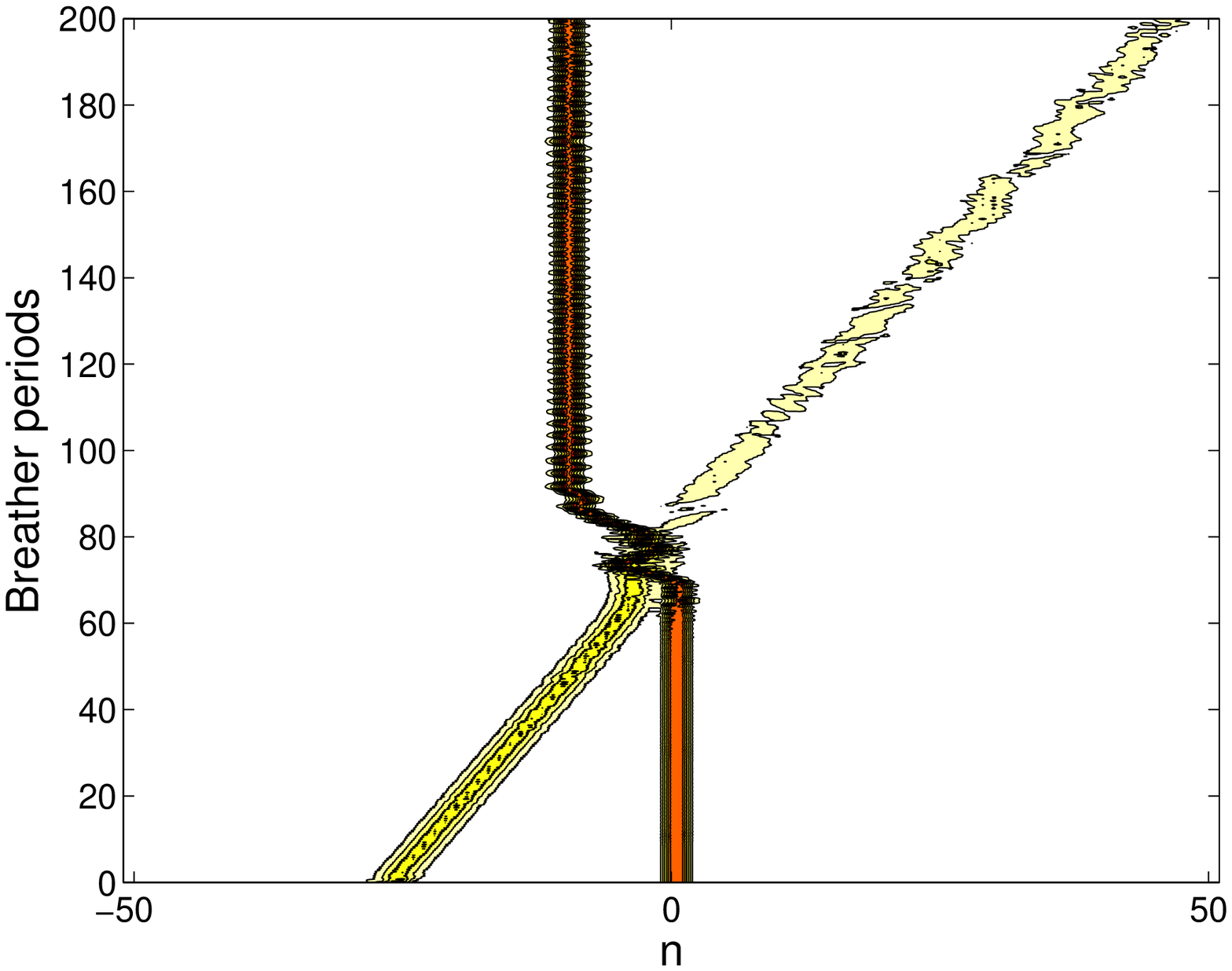} &
    \includegraphics[width=\triplefig]{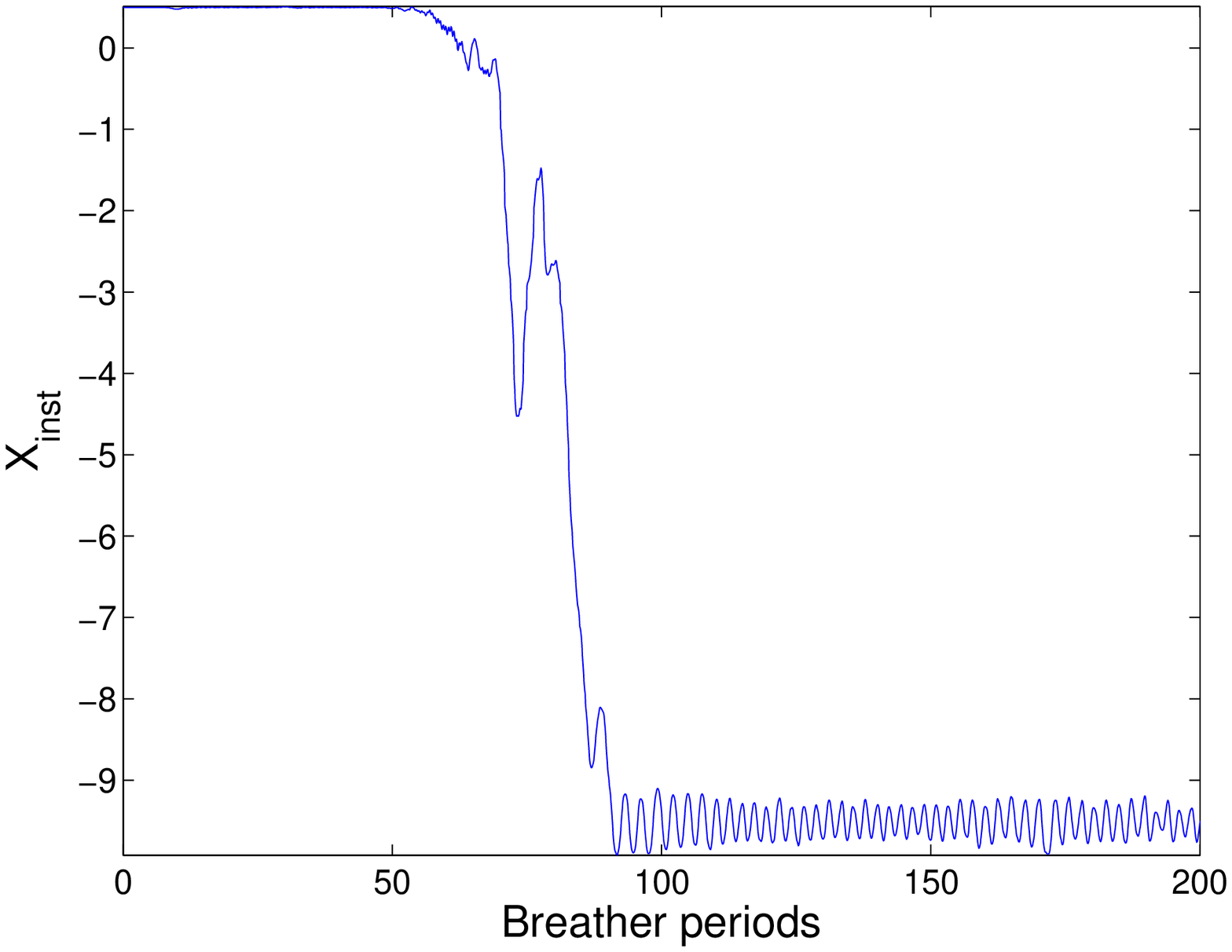} &
    \includegraphics[width=\triplefig]{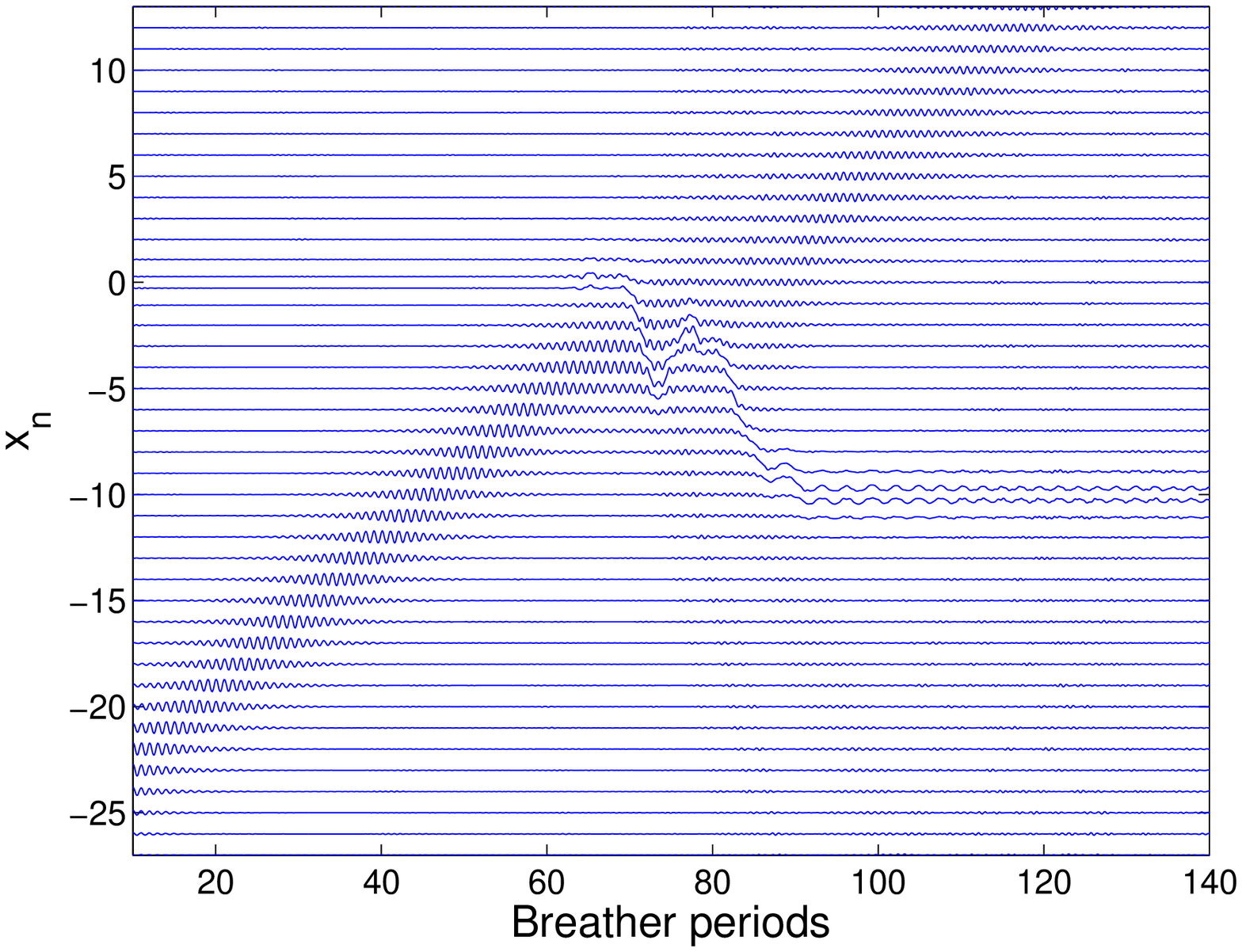} \\
\end{tabular}
\caption{Same as Fig. \ref{fig:edpbk} but for a hopping interstitial
with net backwards displacement. Parameters: $K=0.0050$ and
$b=0.2$.} \label{fig:edpbksw}
\end{center}
\end{figure}

\begin{figure}
\begin{center}
\begin{tabular}{ccc}
    \includegraphics[width=\triplefig]{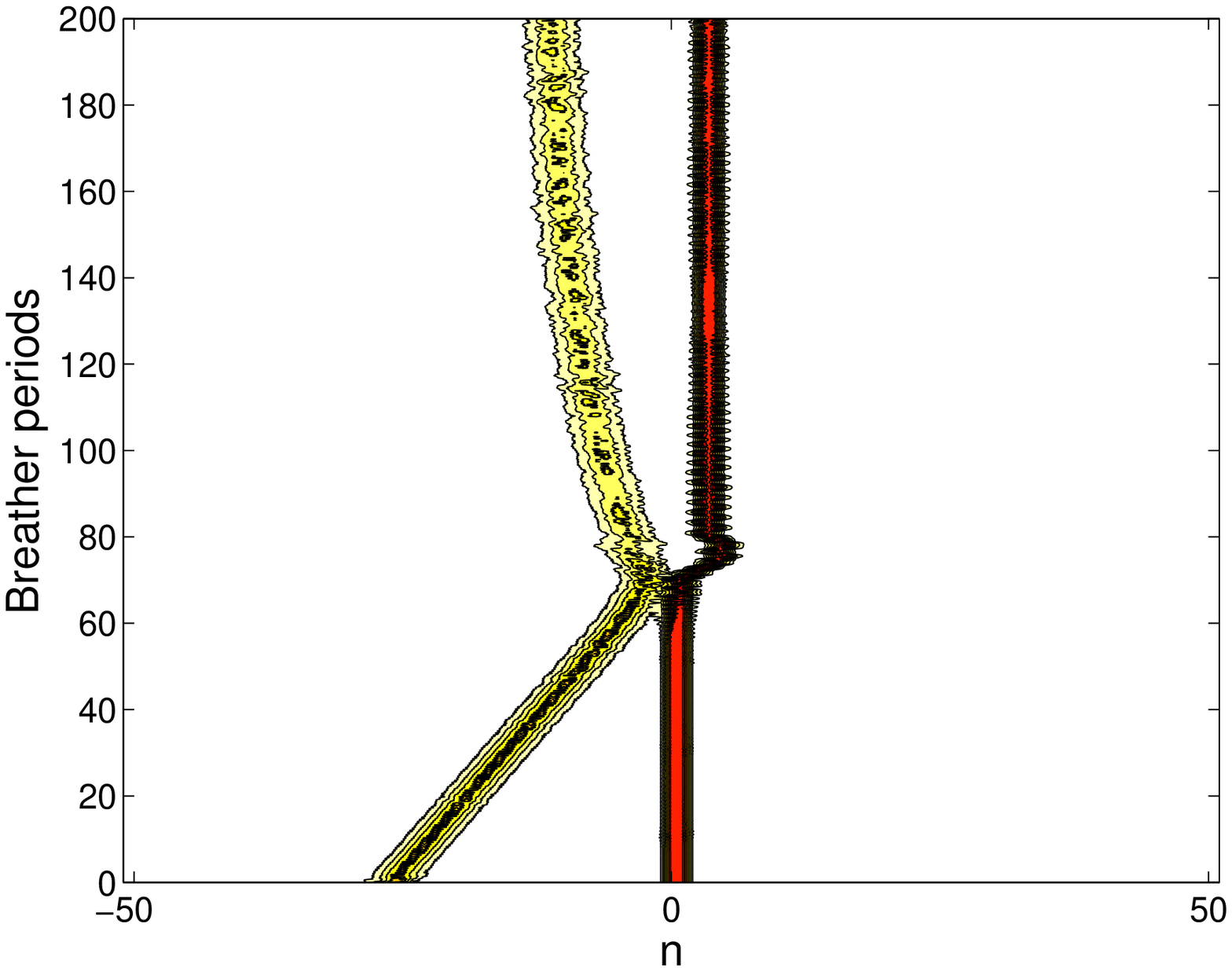} &
    \includegraphics[width=\triplefig]{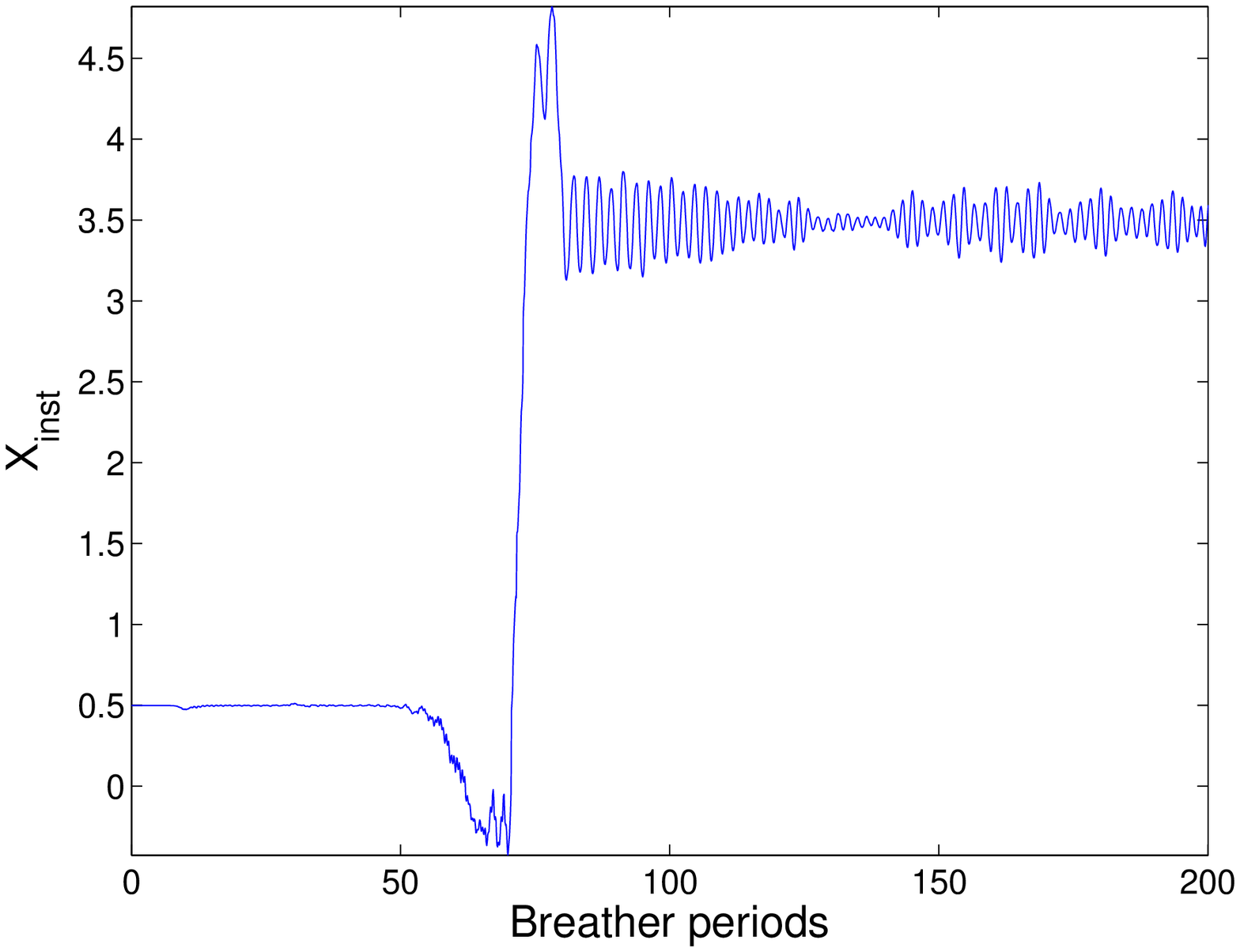} &
    \includegraphics[width=\triplefig]{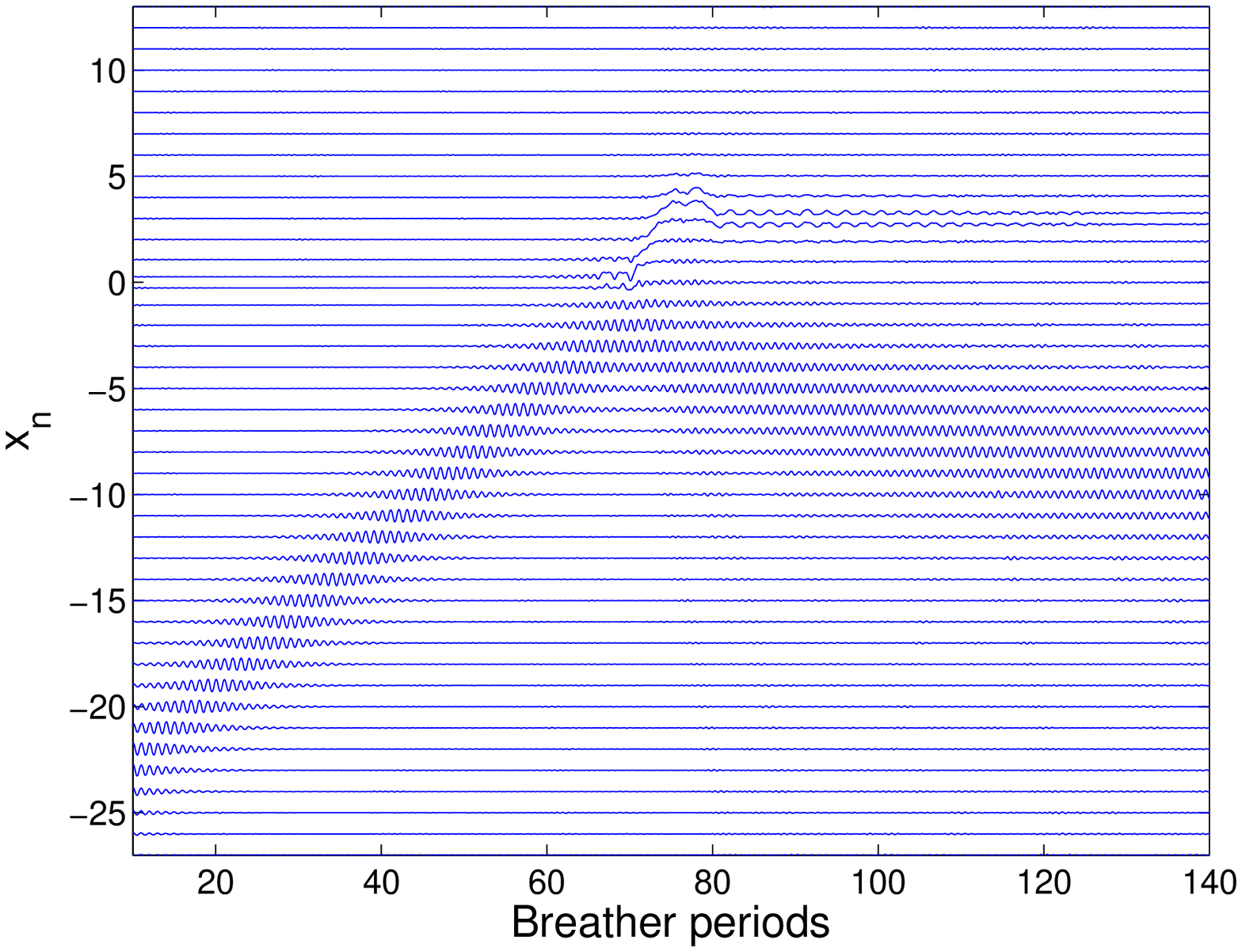} \\
\end{tabular}
\caption{Same as Fig. \ref{fig:edpbk} but for a hopping interstitial
with net forward displacement. Parameters: $K=0.00605$ and $b=0.1$.}
\label{fig:edpfwsw}
\end{center}
\end{figure}

    As a result of the scattering the defect can be put into movement
leading to long--range transport. We have found three
well-differentiated regimes depending on the strength of the
interaction potential. Below a critical value $b\approx 0.83$ the
result of the scattering is unpredictable. The dynamics is extremely
sensitive to initial conditions (value of the perturbation $\lambda$
and  initial position of the breather) and the interstitial can
travel or make random jumps (backward or forward) or even remain at
rest. However, a net backward movement of the defect is only
possible if the interaction potential is strong enough. In fact we
have observed it only for values of $b\lesssim0.69$. An example of a
backwards travelling interstitial is shown in Fig.~\ref{fig:edpbk},
whereas Figs.~\ref{fig:edpbksw} and \ref{fig:edpfwsw} show a
backwards and forwards, respectively, hopping interstitial. In this
case, the interstitial, after several random jumps, remains pinned
on the lattice. These three figures display three panels. Left panel
corresponds to an energy density plot where lines join points with
the same energy in time while darker color indicates larger energy.
Central panel displays the time evolution of the antikink
(interstitial) center of mass. This graph helps to visualize more
clearly the jumps of the interstitial particle and the final
oscillatory state around an equilibrium configuration. Finally,
right panel shows a streak plot with the time evolution of the
breather and the interstitial. It is noteworthy that in our
numerical experiments smaller values of $b$ enhance backward
movement and hopping behaviour of the interstitial particle. This
latest behaviour is the only observed in the harmonic limit of the
interaction potential ($b=0$).

Notice that the complexity of the dynamics is linked to the
discreteness of the F-K model considered \cite{Dmitriev1,Dmitriev2}.
In the continuous limit with $b=0$ the breather-antikink interaction
is an integrable and well-known case, and the resulting scenario is
quite simple: the breather always crosses the antikink which moves
backwards during a brief lapse of time \cite{DEGM}.

\begin{figure}
\begin{center}
    \includegraphics[width=\singlefig]{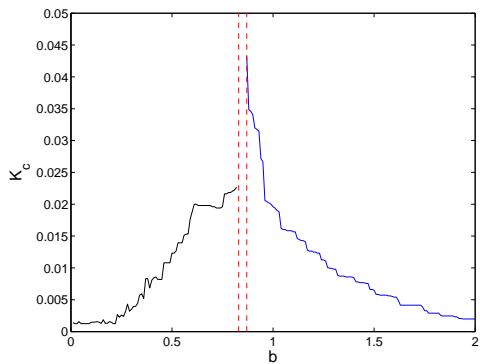}
\caption{Minimum translational energy ($K_c$) of the incoming
breather needed to move an interstitial. In the band $0.83\gtrsim
b\lesssim0.87$ the interstitial always remains pinned on the
lattice.} \label{fig:Kc}
\end{center}
\end{figure}

\begin{figure}
\begin{center}
\begin{tabular}{ccc}
    \includegraphics[width=\triplefig]{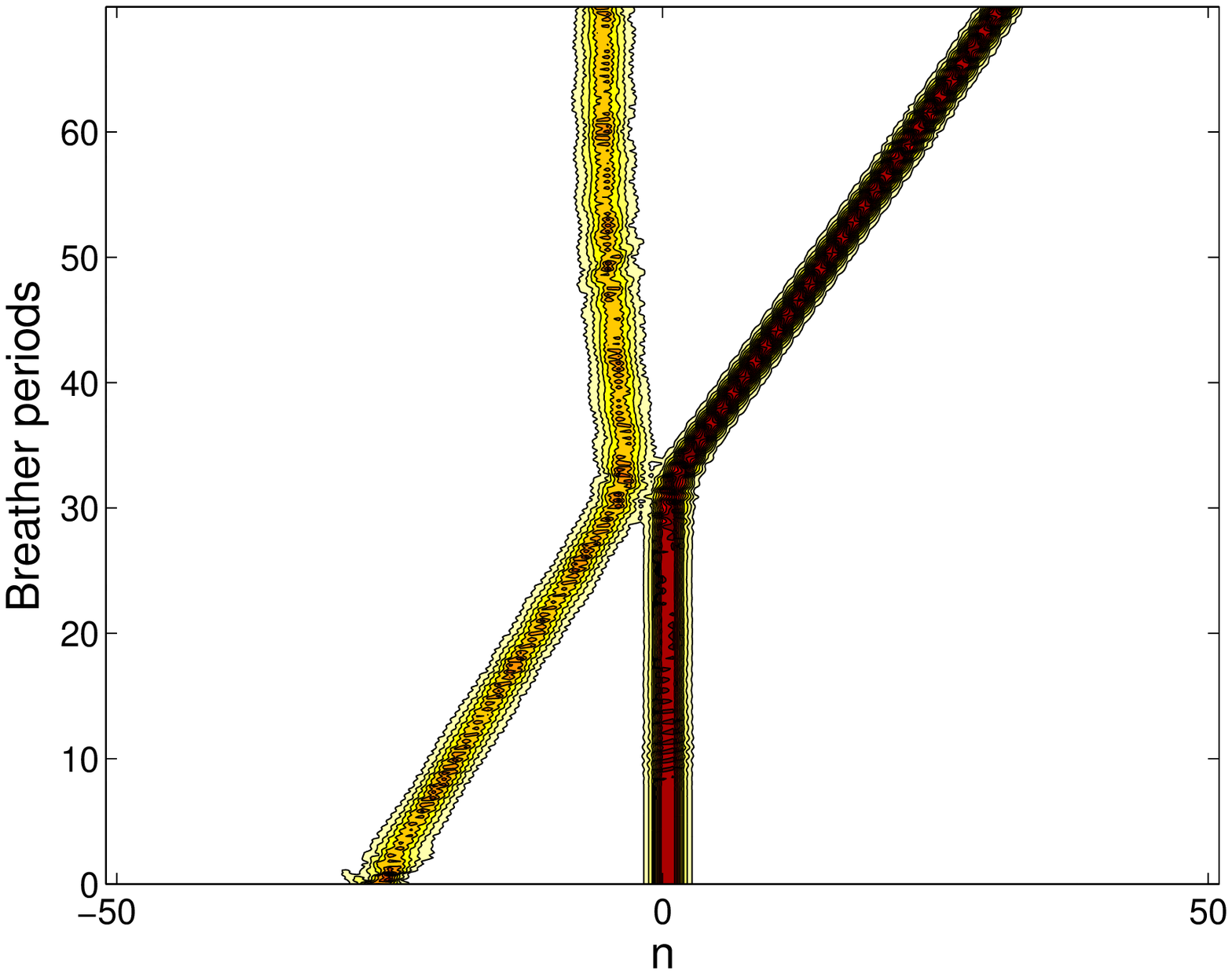} &
    \includegraphics[width=\triplefig]{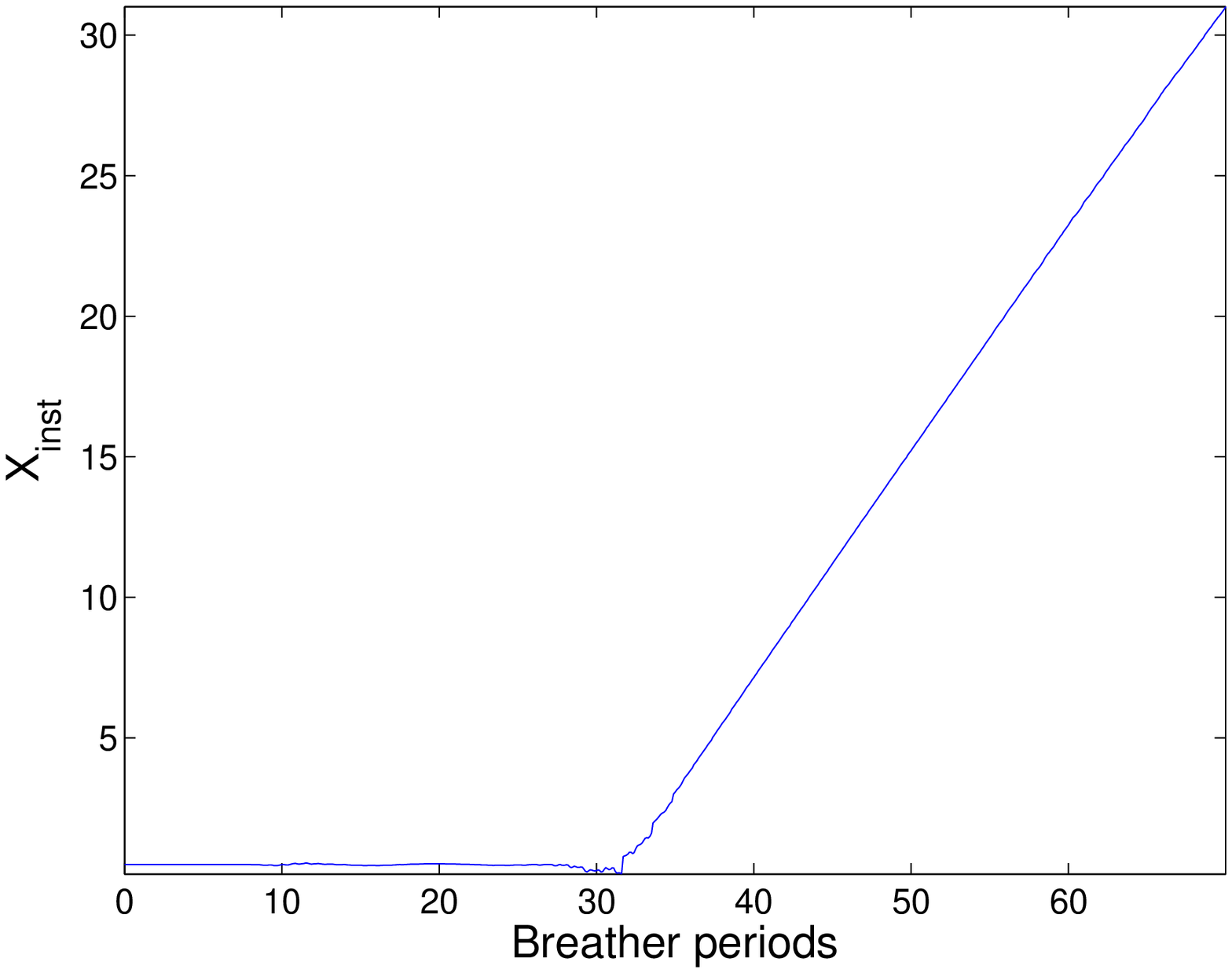} &
    \includegraphics[width=\triplefig]{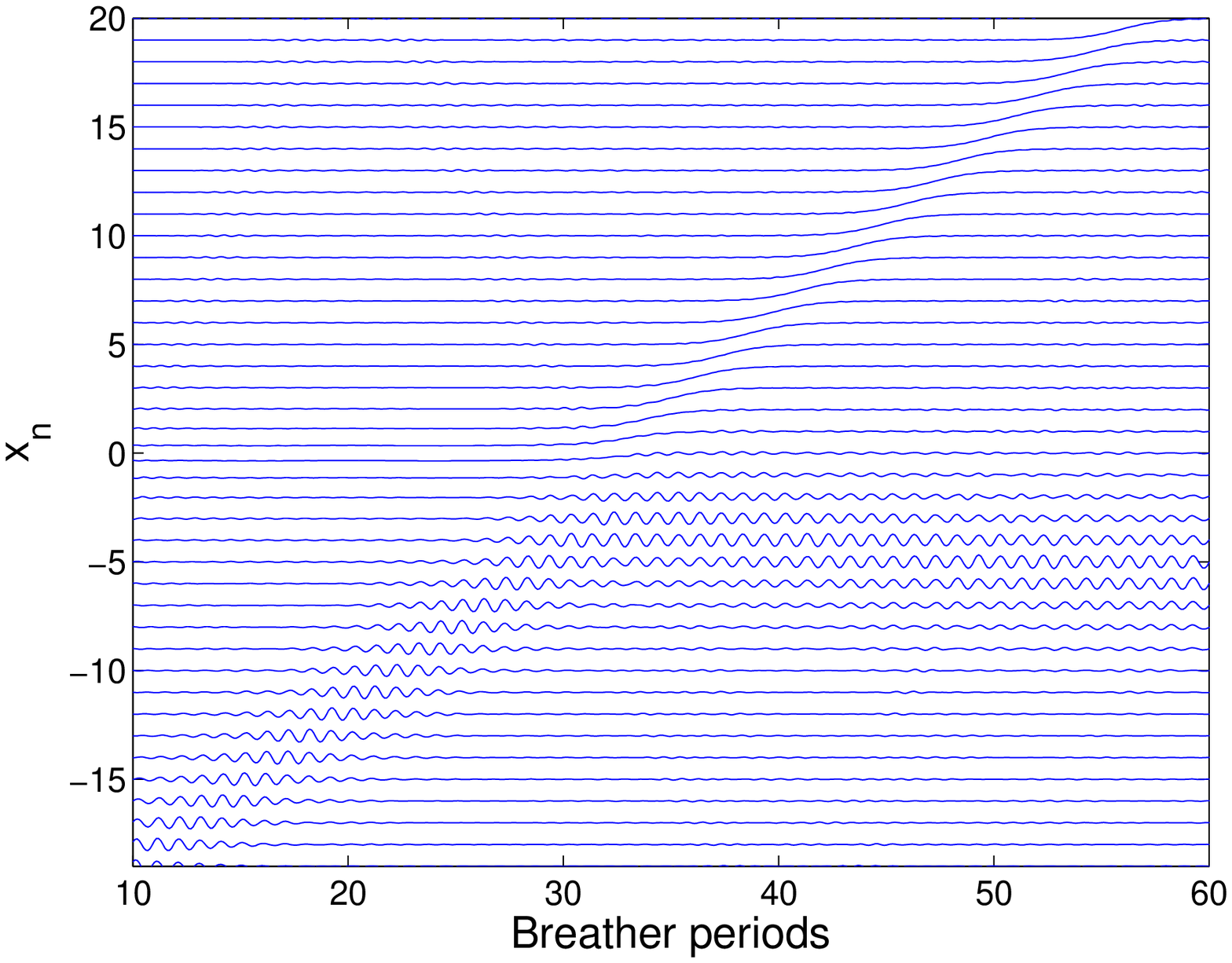} \\
\end{tabular}
\caption{Same as Fig. \ref{fig:edpbk} but for a breather with
$K=0.020$ and $b=1.5$. The interstitial always moves forward with
constant velocity in the parameter region $b\gtrsim0.87$, $K>K_c$ }
\label{fig:edpfw}
\end{center}
\end{figure}

Due to the existence of an activation energy to move an antikink in
the discrete case, interstitial motion is only found above a
threshold value, $K_c$, of the kinetic energy of the incident
breather. In the chaotic regime, $b\lesssim0.83$, this threshold
value, plotted in figure \ref{fig:Kc}, increases monotonically in
contrast with the PNB behavior found in the previous section. On the
contrary, for $b\gtrsim0.87$ we find the opposite tendency: $K_c$
decreases with $b$ indicating a deep change in the dynamics. Indeed
in this parameter regime, for $K>K_c$, the interstitial always moves
forward after the scattering and, remarkably, it always moves with
approximately constant velocity. In this regime, the Morse potential
becomes essentially ``flat'' with a hard core and the dynamics is
dominated by the repulsive part of the interaction potential. An
example can be observed in Fig.~\ref{fig:edpfw}. After the collision
with the breather, interstitial motion is clearly linear in time. Its
velocity has been computed fitting the points of the central panel
with linear regression. In the transition between both regimes, i.e.
for $0.83\lesssim b\lesssim0.87$, the interstitial always remains
pinned on the lattice, at least for those values of $\lambda$ for
which the breather propagates without significant distortion.

\begin{figure}
\begin{center}
\begin{tabular}{ccc}
    \includegraphics[width=\triplefig]{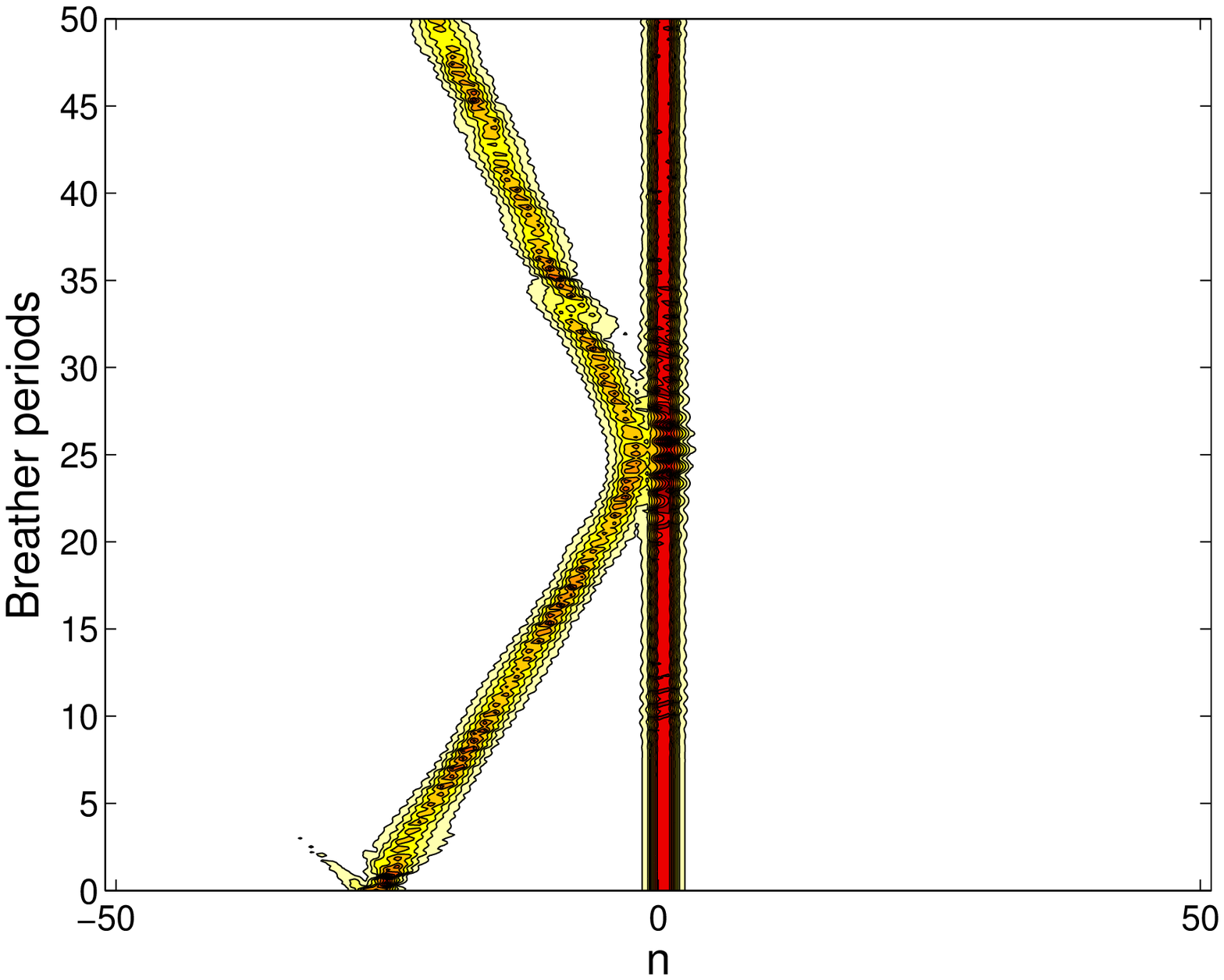} &
    \includegraphics[width=\triplefig]{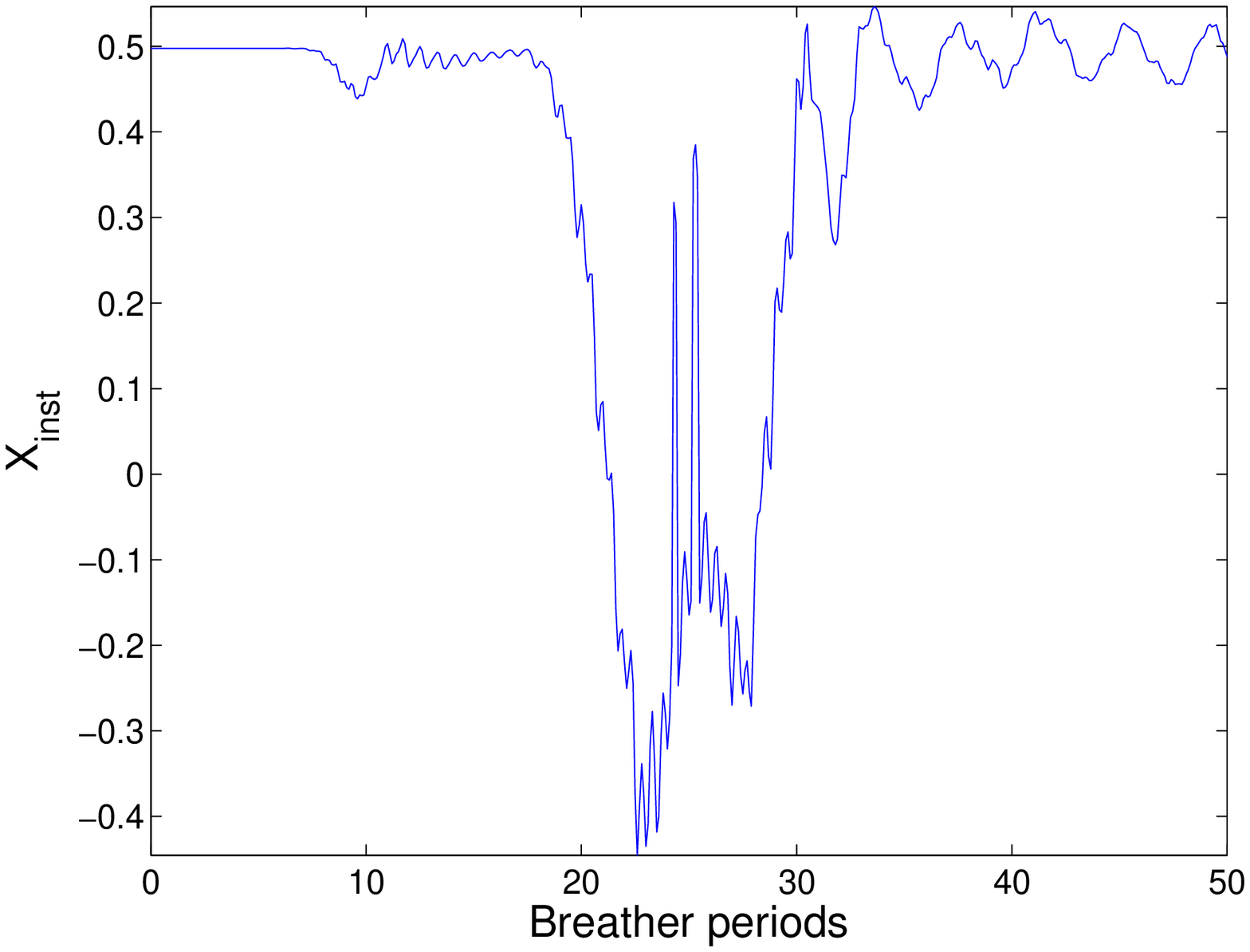} &
    \includegraphics[width=\triplefig]{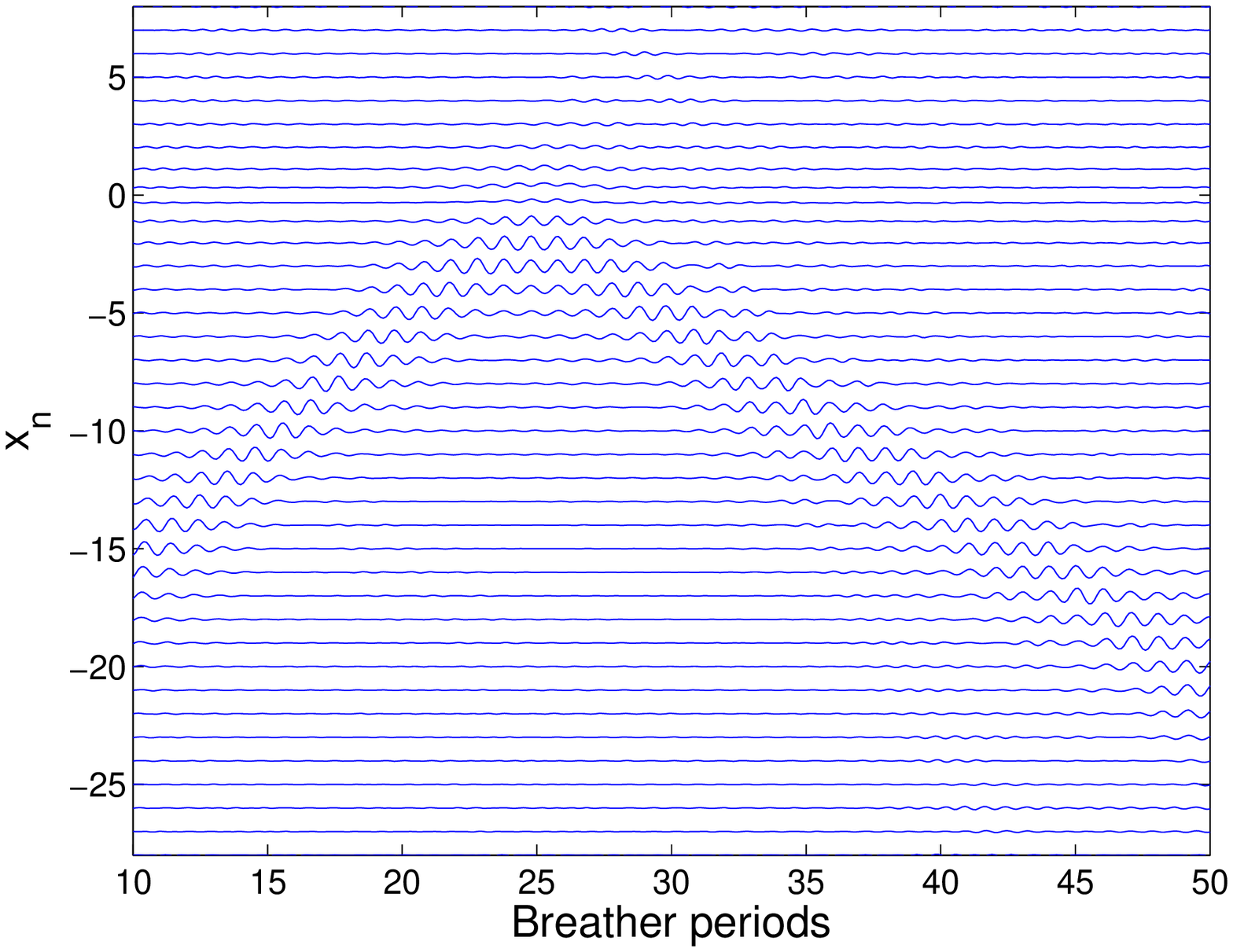} \\
\end{tabular}
\caption{Same as Fig. \ref{fig:edpbk} but for a breather with
$K=0.0162<K_c$ and $b=1$. As the kinetic energy of the incident
breather is below the threshold value $K_c$, the interstitial
remains pinned on the lattice.} \label{fig:edppin}
\end{center}
\end{figure}

\begin{figure}
\begin{center}
    \includegraphics[width=\singlefig]{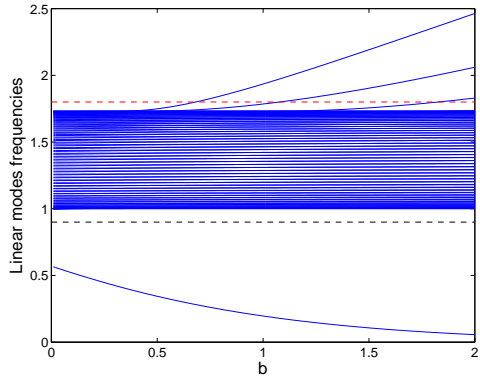}
\caption{Linear modes spectrum of the stable equilibrium
configuration for $C=0.5$. $\wb$ and $2\wb$ are depicted through
dashed lines.} \label{fig:linear}
\end{center}
\end{figure}

Fig.~\ref{fig:edppin} shows the evolution of a pinned interstitial
for $b=1$. In this case the incident breather possesses a
translational energy smaller than the critical value $K_c$ and,
consequently, the interstitial acts as a wall which totally reflects
the breather. It is observed that, after the collision, part of the
breather energy is employed in exciting an internal mode of the
interstitial with a frequency smaller than that of the incident
breather. This linear localized mode corresponds to the line below
the phonon spectrum shown in Fig.~\ref{fig:linear} for the
interstitial stable equilibrium configuration. Note that nonlinear
localized modes do not exist close to the interstitial since the
interaction potential is soft, and the frequency of the linear
localized  mode is always below $\wb=0.9$.

\begin{figure}
\begin{center}
    \includegraphics[width=\singlefig]{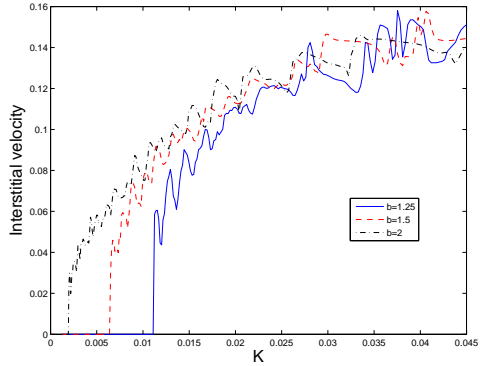}
\caption{Interstitial velocity as a function of the translational
energy $K$ of the incident breather for three different values of
the coupling strength in the regime ($b\gtrsim 0.87$). In this
regime the interstitial always moves forward with constant velocity
because the interaction potential reduces essentially to a repulsive
hard core.} \label{fig:velint}
\end{center}
\end{figure}

As mentioned above, for $b\gtrsim0.87$ a stable interstitial
propagating mode appears if the kinetic energy of the incident
breather is higher than the threshold value $K_c$. In this parameter
region the Morse potential becomes essentially a repulsive
potential. In fact, in the limit $b\rightarrow\infty$ it becomes a
hard-sphere potential. For this reason, in this dynamical regime
interstitial particles move roughly like hard spheres on a wavy
energy landscape. After the breather scattering, the interstitial
particle surmounts the energy barrier of the on-site potential well
and collides with the particle that occupies the following well
transferring its energy and momentum to it. In this way the defect
propagates at constant velocity forever. In figure~\ref{fig:velint}
we have plotted the dependence of the velocity of this propagating
mode on the kinetic energy of the incident breather. One can observe
that just above the threshold energy, interstitial velocity
increases with the kinetic energy, $K$, of the incoming breather.
However for higher values  of $K$ the interstitial velocity tends to
saturate around a value $0.14$, what means that the interstitial
particle moves approximately $0.14\; \frac{2\pi}{\omega_b}\approx 1$
site on the chain per breather period, independently of the coupling
strength.

\begin{figure}
\begin{center}
    \includegraphics[width=\singlefig]{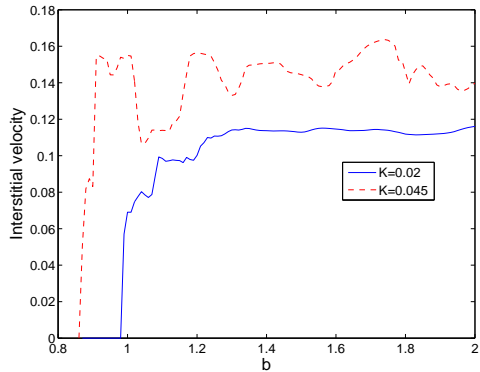}
\caption{Interstitial velocity versus coupling strength for a fixed
translational energy of the incident breather.}
\label{fig:velint2}
\end{center}
\end{figure}

This phenomenon is confirmed in figure~\ref{fig:velint2} where we
have plotted the interstitial velocity versus parameter $b$ for a
fixed value of $K$. Indeed, for $K=0.045$ (dashed line) well above
the energy threshold, interstitial velocity takes roughly the
saturation value $0.14$ independently of the coupling strength.
Intermediate values of kinetic energy as $K=0.02$ (continuous line)
also leads to saturation velocities independently of $b$ but with
values lower than 0.14 and less fluctuations.

\section{Conclusions}
We have presented numerical results arising from the interaction
between a moving discrete breather and an interstitial defect in a
FK chain. The main result is the existence of three differentiated
regimes depending on the strength of the interaction potential. When
the interaction between neighbors is strong the dynamics is chaotic
and the behavior of the interstitial particle is unpredictable: it
can jump backwards, forwards or remains at rest. However, if the
interaction potential is weak enough, the defect moves forwards
along the lattice with constant velocity. This stable propagating
mode had not been observed to our knowledge in previous numerical
studies concerning the interaction between moving breathers and
point defects. The effect is ascribed to the fact that the
interaction potential reduces essentially to a repulsive hard core.
Between these two dynamical regimes there is an narrow intermediate
range of the coupling strength in which the interstitial  always
remains pinned.

Out of that  pinned regime, the kinetic energy of the incoming
breathers must surpass a threshold in order to move the
interstitial. This energy threshold has a non-monotonic behavior. It
grows with parameter $b$ in the chaotic regime, but decreases with
$b$ when the system losses sensitivity to initial conditions and the
propagating mode emerges. With due caution these results can assist
in understanding the interaction of mobile discrete breathers with
true initially stationary interstitial atoms lying adjacent to a
chain in a crystal, which may be of the same or different species
from that of the chain. The experiments reported in Refs.\
\cite{SAR00} and \cite{AMM06} are of each type. In these experiments
an incident discrete breather must supply both the kinetic energy
and the momentum of the interstitial that is put into motion.
Moreover, once set in motion the interstitial, and thus its
influence on the adjacent chain, is expected to move at the same
speed as the discrete breather, thus carrying the defect as opposed
to repeated sweeping by subsequent discrete breathers, which is less
probable.

    Our results are also in accordance with the experimental fact that
interstitial defects diffuse easily and faster than vacancy ones,
and support the hypothesis that scattering with high energy mobile
breathers may play an important role for defect diffusion in
crystals under ion bombardment.

\section*{Acknowledgements}

Two of the authors (JC and BSR) acknowledge sponsorship by the
Ministerio de Ciencia e Innovaci\'on (Spain), project
FIS2008-04848.

\medskip
Received xxxx 20xx; revised xxxx 20xx.
\medskip

\end{document}